# Thermoelectricity of moiré heavy fermions in MoTe$_2$/WSe$_2$ bilayers

Yichi Zhang[1*], Wenjin Zhao[2*], Zhongdong Han[1], Kenji Watanabe[3], Takashi Taniguchi[3], Jie Shan[1,2,4,5†], Kin Fai Mak[1,2,4,5†]

[1]Laboratory of Atomic and Solid State Physics, Cornell University, Ithaca, NY, USA
[2]Kavli Institute at Cornell for Nanoscale Science, Ithaca, NY, USA
[3]National Institute for Materials Science, Tsukuba, Japan
[4]School of Applied and Engineering Physics, Cornell University, Ithaca, NY, USA
[5]Max Planck Institute for the Structure and Dynamics of Matter, Hamburg, Germany

[*]These authors contributed equally
[†]Email: jie.shan@mpsd.mpg.de; kin-fai.mak@mpsd.mpg.de

**Abstract**
Tunable Kondo lattice and heavy fermion physics have been recently reported in moiré materials, but most of the studies have focused on the electrical and magnetic properties. Quantitative thermoelectric measurements, which can reveal entropic information of the heavy fermions, have yet to be achieved. Here, we report a comprehensive thermoelectric study on the moiré heavy fermion phase realized in hole-doped angle-aligned MoTe$_2$/WSe$_2$ bilayers. By electrically gating the material to the Kondo lattice region of the phase diagram, we observe a sign change in the Seebeck coefficient near the Kondo coherence temperature, where the heavy fermion phase with an electron-like Fermi surface evolves into an itinerant Fermi liquid with a hole-like Fermi surface. We compare the results with the semiclassical Mott relation and discuss the observed discrepancies. In addition to the thermal dissociation of Kondo singlets in the heavy Fermi liquid, a sign change accompanied by a strong peak in the Seebeck coefficient is also observed near a Zeeman breakdown of the Kondo singlets, signaling an entropy accumulation. Our results provide entropic information on both the formation and breakdown of heavy fermions in moiré semiconductors.

**Main**
Recently, tunable Kondo lattice and heavy fermion physics have been reported in both graphene[1-3] and transition metal dichalcogenide (TMD)[4-7] moiré materials. Most of the experimental studies to date have focused on the electrical and magnetic properties. Although photo-thermoelectric evidence of topological heavy fermions in twisted bilayer graphene has been reported recently[2,3], quantitative thermoelectric measurements of the heavy fermion phase are still missing because of the difficulty in accurate temperature calibration under a driven thermal current. Accurate measurements of the thermoelectric coefficients are rewarding because they can quantify the entropy per particle (the Seebeck coefficient) and/or per vortex (the Nernst coefficient) in the zero-temperature-gradient limit in strongly correlated materials[8-16] (e.g., cuprate high-temperature superconductors[17,18] and heavy fermion materials[19-21]). In heavy fermion materials, the large Seebeck coefficients due to the heavy electron mass and their temperature and magnetic field dependences can provide invaluable entropic information for the formation of Kondo singlets[22-24]. Performing quantitative thermoelectric measurements on moiré heavy fermion systems is therefore fundamentally important.

In this study, we report measurements of the Seebeck coefficient ($S$) in the Kondo lattice regime of hole-doped angle-aligned MoTe$_2$/WSe$_2$ moiré bilayers. We observe a thermal crossover from an electron-like response ($S < 0$) in the heavy fermion phase to a hole-like response ($S > 0$) at temperatures above the Kondo coherence temperature ($T^*$), where Kondo singlets are dissociated by thermal excitations. The result violates a simple application of the semiclassical Mott relation for $S$ in the heavy fermion phase; possible origins of the violation are discussed. A similar sign change in $S$ and a strong enhancement are also observed near a breakdown of the Kondo singlets by an external Zeeman field at low temperatures. These results are contrasted to the normal behavior of an itinerant Fermi liquid when the Kondo lattice is quenched by gating.

**Moiré Kondo lattice in MoTe$_2$/WSe$_2$ bilayers**
We examine dual-gated Hall bar devices of angle-aligned MoTe$_2$/WSe$_2$ moiré bilayers (Fig. 1a and 1b), in which Kondo lattice and heavy fermion physics have been reported[4-7]. The dual-gated structure allows independent control of the moiré lattice filling factor ($\nu$) and the perpendicular electric field ($E$). The 7% lattice mismatch between the two TMD layers creates a moiré lattice of period $a_M \approx 5$nm and a moiré density $n_M \approx 5 \times 10^{12}$ cm$^{-2}$ (Ref. [4-7,25]). The moiré lattice creates flat bands in both the Mo- and W-layers (we denote MoTe$_2$ and WSe$_2$ by Mo and W, respectively, from now on)[26-31]. At low E-fields, where the topmost valence band of the Mo-layer is higher in energy than that of the W-layer, hole-doping the material depletes electronic states only in the Mo-layer[27,28]; a triangular lattice Mott insulator emerges at half-band-filling (i.e. $\nu = 1$) due to the strong on-site Coulomb repulsion in the Mo-layer[32-35]; the Mo-band is split into the lower and upper Hubbard bands. The E-field tunes the energy difference between the Mo- and W-bands through the interlayer Stark effect[26]. At sufficiently high E-fields, the W-band can be tuned to locate in between the Mo-Hubbard bands[4-7] (Fig. 1c). Doping beyond $\nu = 1$ injects holes only in the W-layer, which are nearly itinerant because of the significantly more dispersive W-band (and thus much weaker correlation) than its Mo-counterpart[27]. The system in this regime can be mapped to a moiré Kondo lattice model with a Kondo exchange interaction ($J_K$) between the local magnetic moments in the Mo-layer and the itinerant holes in the W-layer[27-31,36,37].

The schematic electrostatics phase diagram of the material is shown in Fig. 1c. The Kondo lattice region with doping $\nu = \nu_{Mo} + \nu_W = 1 + \nu_W$ is shaded in blue ($\nu_{Mo}$ and $\nu_W$ denote the filling factor in the Mo- and W-layer, respectively). The E-field span in this region is proportional to the Mott gap $\approx 37$meV in the Mo-layer [see Supplementary Information (SI)]. A heavy Fermi liquid with a large Fermi surface has been reported at temperatures $T \lesssim T^*$; $T^*$ increases with $\nu_W$ as more itinerant holes are available to Kondo-screen the local moments[4]. In addition to thermal dissociation of the Kondo singlets at $T \gtrsim T^*$, an external Zeeman field ($B$) with Zeeman energy $g\mu_B B \gtrsim k_B T^*$ can also break down the Kondo singlets and destroy the heavy Fermi liquid[38] ($g$, $\mu_B$ and $k_B$ denote the electron g-factor, the Bohr magneton and the Boltzmann constant, respectively). This Zeeman breakdown has enabled us to visualize the $\nu = 1 + \nu_W$ Kondo lattice region in the resistivity ($\rho_{xx}$) map under a perpendicular magnetic field $B = 12$T (Fig. 1d); specifically, we observe vertical stripes in $\rho_{xx}$ arising from the quantized Landau levels of the itinerant holes in the W-layer. (See Fig. S1 in SI for additional $\rho_{xx}$ maps at other B-fields.) Similar stripes in $\rho_{xx}$ are also observed in the $\nu = 2 + \nu_W$ region of the phase diagram. Here the topmost moiré valence band of the Mo-layer is fully filled; the lattice of local moments is quenched; the physics is controlled by the itinerant holes remaining in the W-layer. We will use this region as a control experiment in the following discussions. The above results are fully consistent with earlier reports[4-7].

**Thermoelectric measurements**

To quantitatively measure the Seebeck coefficient $S$ of MoTe$_2$/WSe$_2$, we evaporated a platinum heater and encapsulated it between the top and bottom hexagonal boron nitride (hBN) gate dielectrics (Fig. 1a)[13-15]. The high in-plane thermal conductivity of hBN generates a nearly linear temperature gradient across the sample when the heater is turned on[13]; this is confirmed by both finite element analysis (Fig. S2 in SI) and local temperature measurements (inset of Fig. 1e and Fig. S3-7 in SI).

We measured the Seebeck coefficient $S$ by the established $2\omega$ method (Fig. 1b)[13-16,39,40]. We biased an AC current ($I_{ac}$) at frequency $\omega = 11.137$Hz in the heater, which induces a $2\omega$ thermoelectric voltage drop ($V_S$) in the longitudinal direction of the device. To calibrate the local temperature rise ($\Delta T$) induced by heating, we first measured the T-dependence of the two-terminal resistance (heater off) for each transverse electrode pair in the device near the $\nu = 1$ insulating state, where a strong T-dependence is observed; we then measured the heating-induced resistance change at $2\omega$ to obtain $\Delta T$. (Note that gate voltages were chosen such that the Seebeck and Nernst responses are negligible compared to the heating-induced resistance change.) For small $\omega$, $\Delta T$ is determined by the thermal conductance of the substrate (not the sample) and is thus independent of gating[13-15]. The calibrated $\Delta T$ shows a linear dependence on the distance ($L$) from the heater (inset of Fig. 1e); the resultant temperature gradient $\frac{\Delta T}{\Delta L}$ decays quickly with $T$ (Fig. 1e). The Seebeck coefficient is $S = \frac{\Delta V_S/\Delta L}{\Delta T/\Delta L}$ ($\Delta L$ is the distance between two adjacent probe electrodes, see Fig. 1b).

We performed several tests to ensure the reliability of the Seebeck measurement. First, the thermoelectric measurements were performed in the linear response regime. This is illustrated by the linear dependence of $V_S$ on the heater power in Fig. 1f. (See Fig. S9 in SI for additional power dependence measurements.) The temperature difference between adjacent electrodes in the longitudinal direction is also very small compared to the sample temperature $T$ (Fig. 1e). Second, we performed additional grounding tests and frequency dependence measurements to rule out measurement artifacts (Fig. S8 in SI); we also ensured a 90° phase shift in $V_S$ relative to $I_{ac}$ in all measurements. Third, we confirmed the reliability of our Seebeck measurements in the quantum Hall regime under high B-fields. The measured Seebeck coefficient is in good agreement with the expected quantized value $S = \frac{(k_B/e)\ln 2}{n+1/2} \approx \frac{59.7}{n+1/2}\mu$V/K at the $n^{\text{th}}$ quantum Hall state (Fig. S10 in SI)[41-43]. ($e$ is the electron charge.) All these demonstrate the reliability of our measurements. See SI for details on device fabrication, measurements, and simulations.

**Thermoelectric response of moiré heavy fermions**

Figure 2a and 2b show the dependence of $S$ and $\rho_{xy}$, respectively, on $\nu$ and $E$ at $T = 2$K ($\rho_{xy}$ is the Hall resistivity measured at $B = 2$T). The regions of interest, $\nu = 1 + \nu_W$ and $\nu = 2 + \nu_W$, are labeled according to the high B-field map of $\rho_{xx}$ in Fig. 1d. We observe opposite signs for $S$ and $\rho_{xy}$ in both the $\nu = 1 + \nu_W$ and $\nu = 2 + \nu_W$ regions; specifically, $S > 0$ and $\rho_{xy} < 0$, corresponding to hole-like transport, and $S < 0$ and $\rho_{xy} > 0$, corresponding to electron-like transport, are observed at $\nu = 2 + \nu_W$ and $\nu = 1 + \nu_W$, respectively.

We compare the measured Seebeck coefficient $S$ to a simple application of the semiclassical Mott relation for metals $S_{\text{Mott}} = -\frac{\pi^2 k_B^2 T}{3e} \frac{d \ln \sigma_{xx}}{d\epsilon}\Big|_{\epsilon=\epsilon_F} \approx -\frac{\pi^2 k_B^2 T}{3e} \frac{d \ln \sigma_{xx}}{d\nu} \frac{N(\epsilon_F)}{n_M}$ (Ref. [44,45]). Here, $\sigma_{xx} = \frac{1}{\rho_{xx}}$ is the sample conductivity, $\epsilon_F$ is the electron Fermi energy, and $N(\epsilon_F) = \frac{m^*}{\pi \hbar}$ is the density of state at $\epsilon_F$ for the W-layer with an effective hole mass $m^* \approx 0.5 m_0$ obtained from T-dependent Shubnikov-de Haas measurements ($m_0$ is the free electron mass)[4]. Figure 2c shows the calculated map for $S_{\text{Mott}}$; representative linecuts along the center of the $\nu = 1 + \nu_W$ and $\nu = 2 + \nu_W$ regions are shown in Fig. 2d. Whereas a good agreement between $S$ and $S_{\text{Mott}}$ is observed at $\nu = 2 + \nu_W$, the two show opposite signs in most of the regions at $\nu = 1 + \nu_W$. In other words, while $S_{\text{Mott}}$ predicts hole-like transport in the W-layer for both $\nu = 1 + \nu_W$ and $\nu = 2 + \nu_W$, the measured Seebeck coefficient shows electron-like transport in the $\nu = 1 + \nu_W$ region.

To better understand the origin of the disagreement between $S$ and $S_{\text{Mott}}$, we perform T-dependence studies at both $\nu = 1 + \nu_W$ and $\nu = 2 + \nu_W$. We first examine the T-dependence of $S$ at different fillings for $\nu = 2 + \nu_W$ in Fig. 3a and 3b ($E = 0.5$ V/nm). We also compare the results to the metallic T-dependence of $\rho_{xx}$ at the same fillings in Fig. 3c. The Seebeck coefficient $S$ shows a T-linear dependence at low-T and a decreasing slope with increasing $\nu_W$. This T-dependence is well described by the Mott relation for itinerant holes in the W-layer, which predicts $S = \frac{\pi^2}{2} \frac{k_B}{e} \frac{T}{T_F} \approx 425 \frac{T}{T_F} \mu$V/K for impurity-dominated scattering at low-T (Ref. [46]). Here, $T_F \equiv \frac{\epsilon_F}{k_B} = \frac{\hbar^2 \pi n_M \nu_W}{m^* k_B}$ is the Fermi temperature of the itinerant holes. Fitting the data to the Mott relation gives an effective mass $\frac{m^*}{m_0}$ (inset of Fig. 3b) in good agreement with that obtained from T-dependent Shubnikov-de Haas measurements[4]. The deviation from the T-linear dependence at $T \gtrsim 10$K is possibly caused by phonon drag[47,48] and/or flat band effects (SI and Fig. S12).

Next, we study the T-dependence at $\nu = 1 + \nu_W$ in Fig. 3d-f ($E = 0.68$ V/nm). The Seebeck coefficient $S$ scales linearly with $T$ at low T; the absolute slope increases with increasing $\nu_W$; it also changes sign at a characteristic temperature $T^*$ that increases with $\nu_W$. This contrasts with that at $\nu = 2 + \nu_W$, where no sign change is observed. We mark the location of $T^*$ in the T-dependence of $\rho_{xx}$ in Fig. 3f. Interestingly, it corresponds to the previously reported Kondo coherence temperature, where a peak/bump in $\rho_{xx}$ is observed and below which coherent Fermi liquid transport emerges[4-7]. The results strongly suggest the emergence of heavy fermions at $T \lesssim T^*$ as the origin of the observed behavior of the Seebeck coefficient $S$ and its disagreement with $S_{\text{Mott}}$ at $\nu = 1 + \nu_W$.

In a Kondo lattice, the coherent scattering of itinerant electrons with a lattice of local magnetic moments creates Kondo singlets and electrons with a renormalized heavy mass at $T \lesssim T^*$ (Ref. [23,49]). In this heavy fermion phase, the local moments are absorbed by the itinerant electrons, producing a heavy Fermi liquid with a large Fermi surface at low-T. In MoTe$_2$/WSe$_2$, the total hole density of the heavy Fermi liquid is therefore $(1 + \nu_W) n_M$ (rather than $\nu_W n_M$), which is equivalent to an electron density $(1 - \nu_W) n_M$ (Ref. [4]). This explains the electron-like Seebeck coefficient ($S < 0$) and Hall response ($\rho_{xy} > 0$) observed at $T \lesssim T^*$. (The extracted Hall density is also in good agreement with $(1 - \nu_W) n_M$, as demonstrated by earlier studies[4].) With increasing temperature beyond the crossover temperature scale $T^*$, the Kondo singlets are gradually dissociated by thermal excitations; the local moments are released[10], resulting in a change in size of the Fermi surface from $(1 + v_W)n_M$ to $v_W n_M$, thus a sign change in $S$ at $T \gtrsim T^*$, where a better agreement between $S$ and $S_{\text{Mott}}$ is observed (lower panel in Fig. 2d and Fig. S11). Similar T-dependence is also observed in many Ce-based heavy fermion materials, including CeGe$_2$ (Ref. [50]), Ce$_2$N$_7$ (Ref. [51]), CeCu$_2$Si$_2$ (Ref. [21,52]) and CeAl$_2$ (Ref. [53]), and has been taken as thermoelectric evidence of heavy fermions.

Although the above picture qualitatively explains the observed T-dependence of $S$ at $v = 1 + v_W$, extracting quantitative information remains challenging. For instance, if we insist on using $S_{\text{Mott}} \approx \frac{\pi^2}{2} \frac{k_B}{e} \frac{T}{T_F}$ to describe the T-linear dependence in the heavy fermion phase, we would end up with a decreasing $T_F$ with increasing $v_W$, which would contradict the observed dependence of $T^* \sim T_F$ on $v_W$ (Fig. 3d). (Note that $T^*$ is expected to increase with $v_W$ as more itinerant holes are available to Kondo-screen the local moments[30,31,37].) The breakdown of $S_{\text{Mott}}$ here may not be surprising as the step from $S_{\text{Mott}} = -\frac{\pi^2 k_B^2 T}{3e} \frac{d \ln \sigma_{xx}}{d\epsilon}\Big|_{\epsilon=\epsilon_F}$ to $S_{\text{Mott}} \approx -\frac{\pi^2 k_B^2 T}{3e} \frac{d \ln \sigma_{xx}}{dv} \frac{N(\epsilon_F)}{n_M}$ relies on the assumptions of 1) a rigid band, $\sigma_{xx}(\mu, \epsilon) = \sigma_{xx}(\mu + \epsilon, 0)$ (Ref. [2]), and 2) a density-independent scattering rate ($\mu$ is the electron chemical potential). These assumptions are known to fail in many strongly correlated materials[54], although the original Mott relation $S_{\text{Mott}} = -\frac{\pi^2 k_B^2 T}{3e} \frac{d \ln \sigma_{xx}}{d\epsilon}\Big|_{\epsilon=\epsilon_F}$ remains robust as long as a quasiparticle description of transport is valid. Because we cannot directly evaluate $\frac{d \ln \sigma_{xx}}{d\epsilon}\Big|_{\epsilon=\epsilon_F}$ using experimental data, numerically solving the Anderson/Kondo lattice model is required to quantitatively describe the T-dependence of $S$ in the heavy fermion phase.

**Thermoelectric response near Zeeman breakdown of Kondo singlets**
In addition to the breakdown of Kondo singlets by thermal excitations, the Kondo singlets can also be dissociated by an external B-field or a Zeeman energy $g\mu_B B \gtrsim k_B T^*$ (Ref. [38]). Figure 4a and 4b show the dependence of $\rho_{xy}$ and $S$, respectively, on $v$ and $B$ in the $v = 1 + v_W$ region ($T = 2K$ and $E = 0.68V/nm$). Both $\rho_{xy}$ and $S$ show a sign change with increasing B-field; the sign change boundary is slightly different for the two quantities (solid and dashed lines), but the characteristic B-field increases with increasing $v_W$ in both cases. A Landau fan also emerges immediately beyond the characteristic B-field ($B_c$) for $\rho_{xy}$ (solid line). Interestingly, $B_c$ closely traces the peak position of $S$ in Fig. 4b. The appearance of a Seebeck peak at $B = B_c$ is further illustrated by the selected linecuts at constant fillings in Fig. 4c. The peak broadens with increasing temperature (Fig. 4d). We also found an inverse relationship between $B_c$ and the peak value $S(B_c)$ at $T = 2K$, as demonstrated in Fig. 4e.

As demonstrated by earlier studies[4,5], the Hall density for the heavy Fermi liquid ($\rho_{xy} > 0$) is equal to $-(1 - v_W)n_M$ (the – sign is for electron density); it jumps to a small hole density $v_W n_M$ for the itinerant Fermi liquid ($\rho_{xy} < 0$) at $B > B_c \sim \frac{k_B T^*}{g\mu_B}$, where the holes in the W-layer are decoupled from the local moments in the Mo-layer. Because of the much smaller effective mass in the itinerant Fermi liquid, a Landau fan is immediately observed at $B > B_c$. Moreover, the Hall density jump (with B-field) gets sharper with decreasing T, suggesting that the Zeeman breakdown at $B \approx B_c$ is a quantum phase transition rather than a smooth crossover[4].

This conclusion is further supported by the observed Seebeck peak and its thermal broadening near $B_c$. In a simple Fermi liquid with a single Fermi surface and energy-independent scattering mean free path, $S$ has been shown to provide a powerful measure of the entropy per fermion[8,9,12,55]. Although this is no longer strictly correct when these approximations are relaxed, Behnia et al. have demonstrated a phenomenological scaling law $\frac{S}{C_e} N_A e \sim \pm 1$ for $S$ and $C_e$ (the electronic specific heat capacity) in many heavy fermion materials ($N_A$ is the Avogadro number)[12]. The observed Seebeck peak at $B = B_c$ therefore suggests an enhancement in $C_e \propto N(\epsilon_F)$ or in the manybody electronic density of states $N(\epsilon_F)$ at the Zeeman breakdown, and therefore supports the scenario of a quantum phase transition[24]. This entropic interpretation $S \sim \pm \frac{C_e}{N_A e}$ also provides a natural explanation for the inverse relationship between $B_c \sim \frac{k_B T^*}{g \mu_B} \propto \frac{1}{N(\epsilon_F)}$ and the peak value $S(B_c) \propto C_e \propto N(\epsilon_F)$ observed in Fig. 4e. Note that an enhancement in $C_e$ together with an anomaly in the magnetic susceptibility have been reported in the magnetic Kondo breakdowns of several Ce-based heavy fermion materials[56-59]; the results support that the breakdown is a metamagnetic transition rather than a smooth crossover.

**Conclusion**
In conclusion, we have demonstrated thermoelectric evidence of gate-tunable heavy fermions in angle-aligned WSe$_2$/MoTe$_2$ moiré bilayers. The sign and magnitude of the Seebeck coefficient contain important information on the size of the Fermi surface and the entropy per particle. We have observed the breakdown of the Kondo singlets induced by both thermal excitations and an external Zeeman field. The results have also been reproduced in another device (Fig. S13 in SI). Future thermoelectric measurements down to the mK temperature range should be pursued to further reveal the nature of the Zeeman Kondo breakdown, e.g., the possible existence of a quantum critical point[24] and of a continuous metamagnetic phase transition[38].

**Acknowledgements**
We thank Y. Xia, R. Chaturvedi, P. Nguyen, H. Chakraborti, C. Tschirhart, K. Kang and B. Shen for technical discussions. We also thank A. Georges, Q. Si, D. Călugăru and T. Senthil for theoretical discussions.

**Author contributions**
Y. Z. and W. Z. fabricated the devices. Y. Z., W. Z. performed the electrical and thermoelectrical transport measurement and analyzed the data with the help of Z. H. K. W., and T. T. grew the bulk hBN crystals. Y. Z., W. Z., J. S., and K. F. M. designed the scientific objectives and oversaw the project. All authors discussed the results and commented on the manuscript.

# Figures

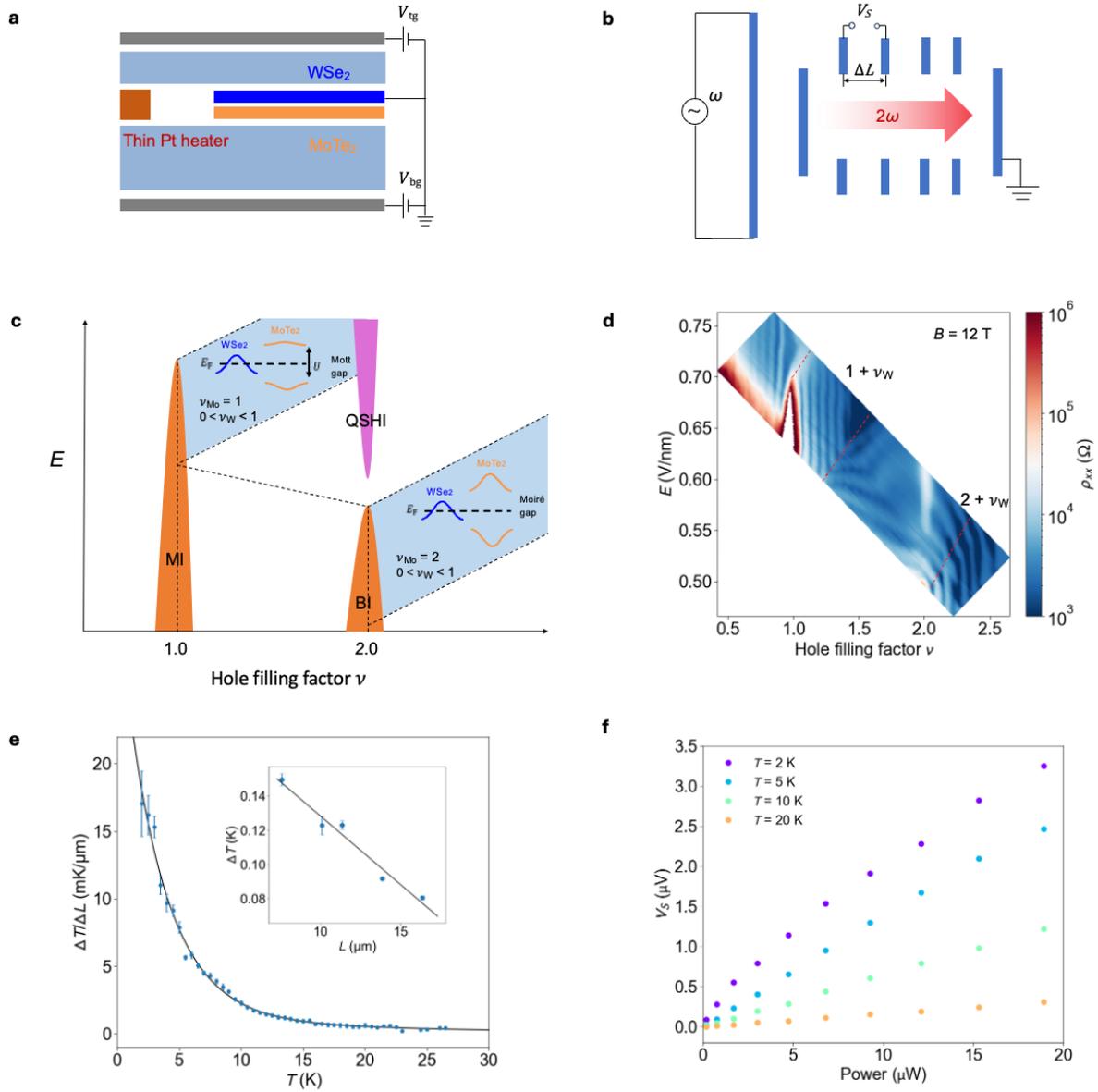

**Figure 1. Thermoelectric measurements on gate-tunable moiré Kondo lattice. a,** Schematic dual-gated angle-aligned MoTe2/WSe2 moiré bilayer device. $V_{tg}$ and $V_{bg}$ denote the top and bottom gate voltages, respectively, applied to the few-layer graphite gate electrodes (grey). The blue areas label the hBN gate dielectrics, which encapsulate the Pt heater. **b,** Schematic $2\omega$ method for measurements of the Seebeck coefficient. An AC excitation current at frequency $\omega$ = 11.137Hz is applied to the heater, which generates a temperature gradient and a Seebeck voltage $V_S$ at frequency $2\omega$ (arrow). $\Delta L$ is the distance between adjacent probe electrodes. **c,** Schematic phase diagram of MoTe2/WSe2 in $\nu$ and $E$. MI, BI and QSHI denote a Mott insulator, a band insulator and a quantum spin Hall insulator. The upper blue shaded region with $\nu = \nu_{Mo} + \nu_W = 1 + \nu_W$ is the Kondo lattice region. $U$ and $E_F$ denote the on-site Coulomb repulsion and the Fermi level (dashed line), respectively. Blue and orange bands label the moiré band of the W-layer and the lower and upper Hubbard bands of the Mo-layer, respectively. The lower blue shaded region

with $\nu = 2 + \nu_W$ is used as a control experiment. The blue and orange bands here label the moiré bands of the W- and Mo-layer, respectively. **d,** Longitudinal resistivity $\rho_{xx}$ as a function of $\nu$ and $E$ at $T = 2K$ and $B = 12T$. The phase boundaries of the $1 + \nu_W$ and $2 + \nu_W$ regions are identified from the Landau level features and marked by red dashed lines. **e,** Temperature gradient $\frac{\Delta T}{\Delta L}$ between adjacent probe electrodes as a function of $T$ (see main text and SI for details on calibration). The solid line is a fit to the data by $Ae^{-\alpha T} + Be^{-\beta T}$. Inset: Temperature rise at $T = 2K$ as a function of the distance to the heater. The solid line is a linear fit to the data. **f,** Power dependence of $V_S$ measured at $T = 2K, 5K, 10K$ and $20K$. The gate voltages are fixed at $V_{tg} = -3.7V$ and $V_{bg} = 0.5V$. Thermoelectric measurements were performed in the linear response regime.

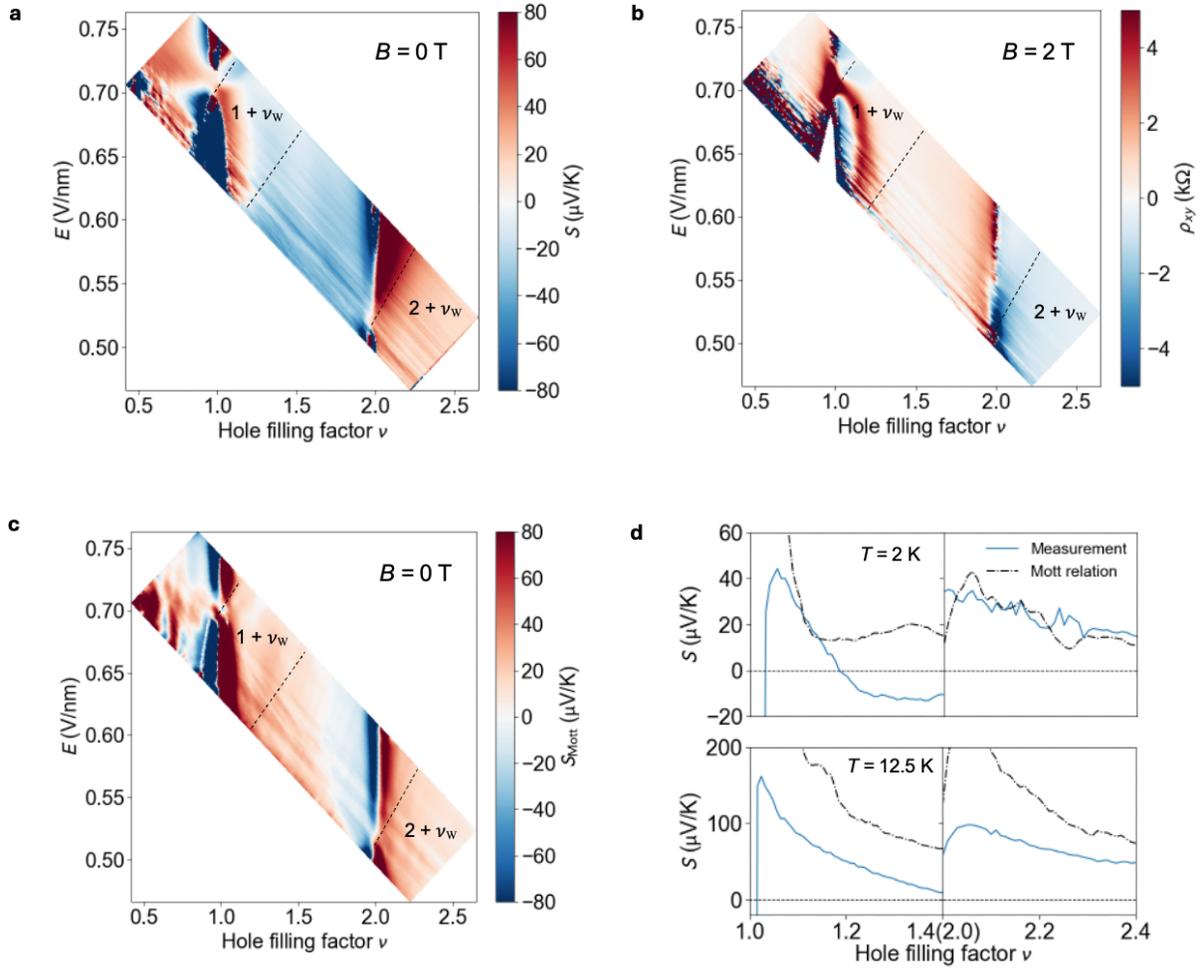

**Figure 2. Seebeck coefficient in a moiré Kondo lattice. a,** Seebeck coefficient $S$ as a function of $\nu$ and $E$ at $T = 2K$ and $B = 0T$. **b,** Hall resistivity $\rho_{xy}$ (measured at $B = 2T$) as a function of $\nu$ and $E$ at $T = 2K$. **c,** The Seebeck coefficient $S_{\text{Mott}}$ obtained using a simple application of the Mott relation as a function of $\nu$ and $E$ (see main text). The dashed lines in **a-c** mark the phase boundaries of the $1 + \nu_W$ and $2 + \nu_W$ regions. **d,** Comparison of the measured Seebeck coefficient $S$ (solid lines) and the calculated $S_{\text{Mott}}$ (solid-dotted lines) at $\nu = 1 + \nu_W$ (left panel, line cut at $E = 0.68V/nm$) and $\nu = 2 + \nu_W$ (right panel, line cut at $E = 0.52V/nm$) for $T = 2K$ (top) and 12.5K (bottom).

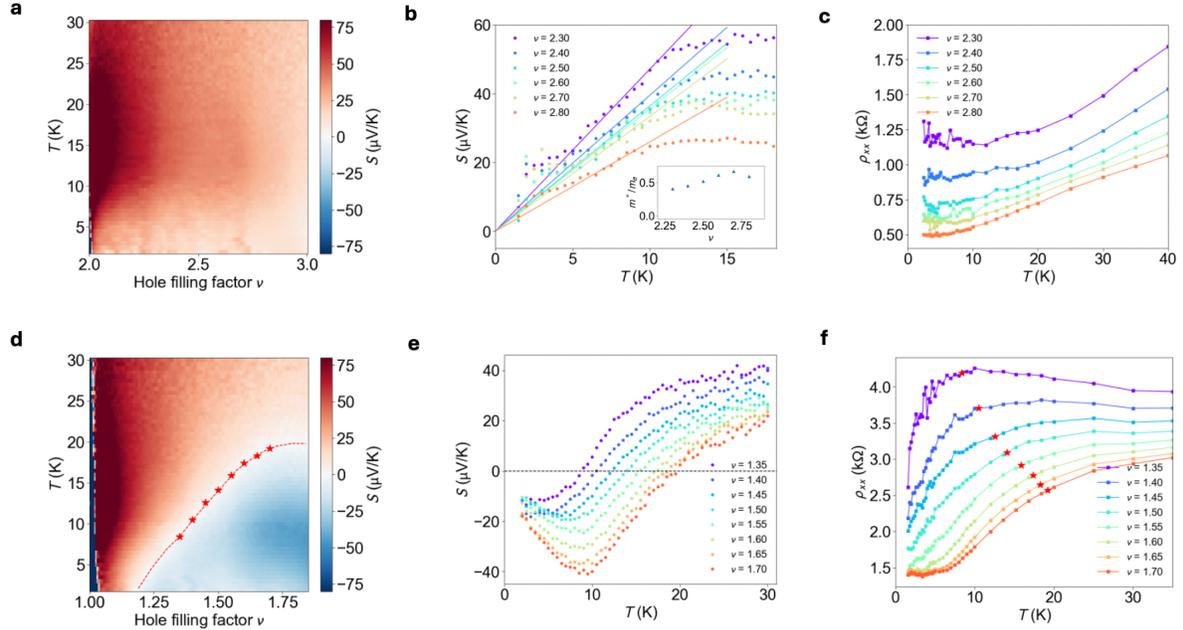

**Figure 3. Temperature-dependent thermoelectric response for heavy fermions. a,d,** Seebeck coefficient $S$ as a function of $\nu$ and $T$ at $E = 0.50\,\text{V/nm}$ (**a**, the $2 + \nu_W$ region) and at $E = 0.68\,\text{V/nm}$ (**d**, the $1 + \nu_W$ region). **b,e,** Selected line cuts at constant fillings from **a** (**b**) and from **d** (**e**). Solid lines in **b** are linear fits to the data in the low-temperature regime. Inset in **b**: Extracted effective mass from the slope of the linear fits as a function of $\nu$. The dashed line in **e** marks the sign change in $S$. **c,f,** Temperature dependence of $\rho_{xx}$ at the same fillings and E-field as **b** (**c**) and as **e** (**f**). The stars in **d** mark the locations for the sign change in $S$, which correlate with the Kondo coherence temperature $T^*$ marked by the same stars in **f**.

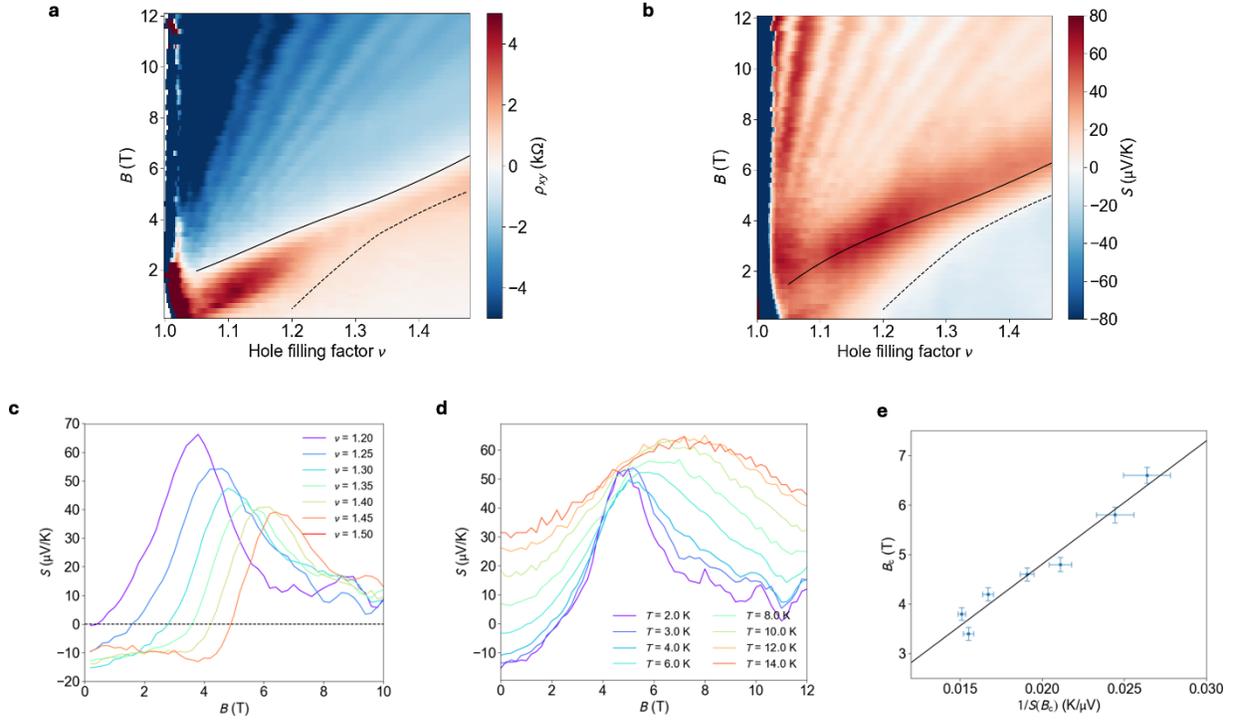

**Figure 4. Thermoelectric response near Zeeman breakdown of Kondo singlets. a,b,** $\rho_{xy}$ (**a**) and $S$ (**b**) as a function of $\nu$ and $B$ at $E = 0.68$V/nm and $T = 2$K. The Zeeman breakdown occurs at the characteristic B-field, $B_c$, for $\rho_{xy}$ (solid lines). The sign change in $S$ (dashed lines) happens at a slightly lower B-field than $B_c$, likely due to particle-hole asymmetry. **c,** Selected line cuts from **b** at constant fillings showing the evolution of the Seebeck peak with filling. **d,** The Seebeck coefficient $S$ as a function of $B$ at varying temperatures ($\nu = 1.3$ and $E = 0.68$V/nm). **e,** Inverse relationship between $B_c$ and the peak value $S(B_c)$ at $T = 2$K. The solid line is a linear fit to the data. The error bar is estimated from the noise level of the Seebeck measurement and the uncertainties in determining the peak position and height of the Seebeck coefficient.

# Supplementary materials for
## "Thermoelectricity of moiré heavy fermions in MoTe₂/WSe₂ bilayers"

Yichi Zhang[*], Wenjin Zhao[*], Zhongdong Han, Kenji Watanabe, Takashi Taniguchi, Jie Shan[†], Kin Fai Mak[†]

[*]These authors contributed equally
[†]Email: jie.shan@mpsd.mpg.de; kin-fai.mak@mpsd.mpg.de

## Methods

### Device fabrication

We used the layer-by-layer dry transfer technique[1,2] to fabricate dual-gated MoTe₂/WSe₂ devices. All devices were transferred onto silicon substrates with pre-patterned electrodes (5nm Ti and 35nm Pt). We first transferred a bottom hBN-graphite gate to the substrate. Pt contacts and heater (8nm thick) were then deposited onto the bottom gate using standard electron beam lithography and evaporation, followed by another step of lithography and evaporation to connect the Pt contacts and heater to the pre-patterned electrodes by 5nm Ti and 40nm Au. To remove the lithographic resist residue, we used an AFM tip to repeatedly scan the channel region in contact mode. The remaining layers (top hBN-graphite gate and the TMD layers) were then picked up and transferred onto the patterned bottom gate in a glove box with O₂ and H₂O levels below 1ppm. The crystallographic orientations of the WSe₂ and MoTe₂ flakes were determined by the second harmonic generation (SHG) technique[3,4]. To distinguish 0° and 60° stacking orders, we cut a portion of the TMD monolayers and aligned them on a separate substrate and performed SHG measurements again. We selected the stacking order with destructive interfering SHG signals.

### Electrical transport measurements

Electrical transport measurements were performed in a closed-cycle ⁴He cryostat (Oxford TeslatronPT). The standard AC lock-in technique ($\omega = 11.137$ Hz) was used to measure the four-terminal resistance at a small source-drain bias voltage $V_b = 1$ mV. A voltage pre-amplifier with 100MΩ impedance was used to amplify the voltage signals at the probe electrodes. The hole density $n$ and electric field $E$ were calculated from a parallel plate capacitor model, $n = \varepsilon_r \varepsilon_0 (\frac{V_{\text{tg}}}{d_{\text{tg}}} + \frac{V_{\text{bg}}}{d_{\text{bg}}})$ and $E = (\frac{V_{\text{bg}}}{d_{\text{bg}}} - \frac{V_{\text{tg}}}{d_{\text{tg}}})/2$, where $V_{\text{tg}}$ and $V_{\text{bg}}$ are the top and bottom gate voltages, $d_{\text{tg}}$ and $d_{\text{bg}}$ are the thicknesses of the top and bottom hBN dielectrics, and $\varepsilon_r$ is the out-of-plane dielectric constant of hBN. The values of $d_{\text{tg}}$ and $d_{\text{bg}}$ were calibrated from the measured Landau level degeneracy under a finite magnetic field. The longitudinal and Hall resistivities ($\rho_{xx}$ and $\rho_{xy}$) were calculated by symmetrizing and anti-symmetrizing the measured resistances at $\pm B$ to remove longitudinal-transverse couplings in the device.

### Mott gap size estimation

The Mott gap size in the Mo-layer is estimated from the E-field span of the $1 + \nu_W$ region. The applied E-field shift the W-band via the Stark effect, $d \times E$, where $d \approx e \times 0.26$ nm is the interlayer dipole moment. We estimated the Mott gap to be 37meV, which is close the value 32meV reported in a previous study[1].

**Finite element simulation of the temperature gradient**

The linear temperature profile across the device was further confirmed by solving the three-dimensional heat diffusion equation by finite element simulations (COMSOL). Figure S2a shows an optical micrograph of a bilayer device. In COMSOL, flakes were modelled by rigid cuboids with thermal contacts (Fig. S2b and c). We also included interfacial thermal resistors $R$ to establish a thermal contact for the unphysical suspended interfaces between the bottom gate hBN and the top gate hBN (or the top graphite gate, see Fig. S2b).

We used real device geometry for our simulations. The thickness of the top (4nm) and bottom (7nm) hBN were determined from the measured Landau level degeneracy. The thickness of the top and bottom graphite gate electrodes (4nm) was estimated from the optical contrast under microscopes. The thickness of the $MoTe_2/WSe_2$ heterobilayer is approximately 2nm. The thickness of the Pt heater is also set to 2 nm for convenience. The silicon wafer is modelled as a 500µm cube with a 275nm thick $SiO_2$ layer on its surface. The bottom of the wafer is connected to an infinite cold reservoir at temperature $T$ (the probe temperature of the cryostat).

The steady state solution was found by solving the heat diffusion equation $\boldsymbol{Q} = \nabla \cdot \boldsymbol{q}$, where $\boldsymbol{Q}$ is the rate at which energy is generated per unit volume and $\boldsymbol{q} = -k\nabla T$ is the heat flux ($k$ is the thermal conductivity). A constant power $P = 4.7\mu W$ was applied to the heater. We followed the simulation of Paul et al.[5] to include both the thermal conductivity $k$ of each layer and the interfacial thermal conductance $G$.

- Thermal conductivity of hBN and graphite
  hBN and graphite have high in-plane thermal conductivity and low out-plane thermal conductivity. We took the graphite thermal conductivity from the COMSOL data library and the hBN thermal conductivity from an earlier experiment[6].
- Thermal conductivity of $MoTe_2/WSe_2$
  We used the reported thermal conductivity data for $WS_2$ (Ref. [7]) to approximate the thermal conductivity of the $MoTe_2/WSe_2$ heterobilayer. Note that, due to the much higher thermal conductivity of hBN and graphite and the much thicker hBN layers, the thermal conductivity of $MoTe_2/WSe_2$ only plays a minor role in determining the temperature gradient.
- Thermal conductivity of Si and $SiO_2$
  The thermal conductivity of silicon depends on the doping. The silicon wafer we used (WaferPro) is p-type with boron dopant. We used the measured thermal conductivity of boron-doped Si (doping density $3\times10^{20}$ atoms/cm$^3$)[8] and the reported thermal conductivity of $SiO_2$ (Ref. [9]) in our simulation.
- Interfacial thermal conductance for the hBN/graphite and TMD/hBN interfaces
  Temperature dependence of the interfacial thermal conductance for hBN/graphite and $MoS_2$/hBN have been measured by Raman spectroscopy[10]. Due to lack of available data, we used the value of hBN/graphite for $SiO_2$/graphite and hBN/hBN (Ref. [5]), and used the value of $MoS_2$/hBN for $WSe_2$/hBN and $MoTe_2$/hBN. Note that although the lack of interfacial thermal conductance data in the literature affects the absolute value of the simulated temperature profile, the overall temperature profile is only weakly influenced by the inaccurate interfacial thermal conductance data.

Figure S2d shows the simulated steady-state temperature field at $T = 2K$. The heater induces a local temperature rise which drops quickly away from the heater. In Fig. S2e, we compare the temperature profile at a line cut through the center of the device for the case with an encapsulated heater and the case with a heater on the silicon substrate. It is found that the high in-plane thermal conductivity of hBN and graphite help to establish an approximately linear temperature drop across the device[5] for the case with an encapsulated heater. Such temperature profile is also confirmed experimentally (inset of Fig. 1e in main text).

We also simulated the temperature difference $\Delta T$ between the two ends of the device as a function of $T$ in Fig. S2f. The simulated $\Delta T$ is largely consistent with our experimental calibration. As mentioned above, the discrepancy in magnitude is likely caused by the lack of thermal conductivity data at the various interfaces in our device structure.

**Temperature gradient calibration**
We used the temperature dependence of the two-terminal resistance to calibrate the local temperature rise induced by the heater. We chose two adjacent electrodes (marked by the yellow star in Fig. S3) with small contact resistance (~10kΩ) for local calibration. First, we applied a constant AC bias $V_{ac} = 1\text{mV}$ ($\omega = 11.137\text{Hz}$) across the electrodes and measured the bias current $I$ as a function of temperature (Fig. S3a). This $I$-$T$ curve was served as a thermometer for the next step. Second, we applied a DC bias voltage $V_{dc} = 1\text{mV}$ across the same electrodes (Fig. S3b) and turned on the heater with an AC heater bias $V_h$. The temperature modulation $\Delta T$ caused by $V_h$ induces a $2\omega$ current modulation $\Delta I$ in the sample, which was measured by a lock-in amplifier at $2\omega$. The local temperature rise is given by $\Delta T = \frac{\Delta I}{dI/dT}$, where $dI/dT$ was extracted from the experimental $I$-$T$ curve obtained from the previous step. Note that $\Delta T$ is a root-mean-square value.

The above calibration process was carried out at a fixed $V_{tg} = -3.525\text{V}$ and at varying $V_{bg}$ from 2.7V to 2.8V close to the $\nu = 1$ insulating state (dashed line in Fig. S4a and b). The line cut was chosen such that the two-terminal resistance is sensitive to temperature while the Seebeck current is negligible compared to the heater-induced $\Delta I$. At $T = 2\text{K}$, the longitudinal resistance is about $10^4 \Omega$ and the Seebeck current is $I_S = V_S/R_{xx} \approx 1\mu\text{V}/10^4\Omega \approx 0.1 \text{ nA}$, which is an order of magnitude smaller than $\Delta I$. Note that the temperature rise $\Delta T$ is largely determined by the thermal conductivity of the substrate and the interfacial thermal conductance; it is independent of $V_{tg}$ and $V_{bg}$ and the specific line cut we chose for calibration.

Here we present the detailed calibration process for the electrode pair labelled by the yellow star in Fig. S3. Fig. S5a shows the sample current $I$ under $V_{ac} = 1\text{mV}$ at the above chosen line cut at different temperatures. The sample is insulating at the line cut, which ensures a strong temperature dependence at low $T$ (Fig. S5b). The extracted slope reaches 2.59nA/K at $T = 2\text{K}$ and decreases to 0.21nA/K at $T = 30\text{K}$ for $V_{bg} = 2.75\text{V}$. We also calculated the temperature dependence of $dI/dT$ for all the back gate voltages along the line cut (Fig. S5c).

The local temperature rise $\Delta T$ decreases almost exponentially as the temperature increases (Fig. S2f), making $\Delta I$ hard to measure at elevated temperatures. To resolve this issue, we measured $\Delta I$ independently at four temperature windows (2.0-5.0K, 5.5-10.0K, 10.5-20.0K and 20.5-30.0K) using different heater voltages ($V_h$ =0.1, 0.2, 0.3 and 0.8V, respectively). The measured $\Delta I$ for each window (Fig. S6a, d, g, j) was converted to $\Delta T$ (Fig. S6b, e, h, k) using the $dI/dT$ data in Fig.

S5c. We then averaged over 100 back gate voltages to obtain the calibrated value of $\Delta T$ shown in Fig. S6c, f, i, l.

The same process was repeated for 5 different electrode pairs along the longitudinal axis of the device (yellow stars in Fig. S2a) to obtain the temperature profile and the temperature gradient. Fig. S7 shows the temperature gradient $\frac{\Delta T}{\Delta L}$ calculated at the two end points of the device separated. by 8.75μm (inset of Fig. S7). The temperature gradient was then normalized to the fixed heater voltage $V_{h0} = 0.1$V (the voltage we used in the Seebeck coefficient measurement) by multiplying the factor $\left(\frac{V_{h0}}{V_h}\right)^2$. The normalized temperature gradient is well fitted by the form $Ae^{-\alpha T} + Be^{-\beta T}$ (solid line in Fig. S7).

**Thermoelectric measurements**
Thermoelectric measurements were performed in the same closed-cycle $^4$He cryostat. An AC voltage $V_h$ of frequency $\omega = 11.137$Hz was applied to the heater; the Seebeck and Nernst voltages were measured at $2\omega$ by the standard lock-in technique. Frequency tests confirmed the applicability of the low-frequency lock-in technique to thermoelectric transport (Fig. S8a). To measure the open-circuit voltages, one end of the sample was grounded to set a reference point. In practice, we found grounding different electrodes may alter the equipotential lines in the device channel, which can result in slightly different thermoelectric voltages. This difference could be significant as the sample is inhomogeneous. For this reason, we tested the Seebeck voltage for different grounding schemes (Fig. S8b), which confirms the consistency with respect to different grounding in our measurements.

As pointed out in the main text, sample heating may affect the low-temperature thermoelectric measurement. To obtain accurate value of the thermoelectric coefficient, it is crucial to perform measurements in the linear-power regime. Power dependence of the Seebeck voltage was measured at $T = 2, 5, 10$ and $20$K at $V_{tg} = -3.7$V and at $V_{bg} = 0 \sim 3.5$V. (Fig. S9a-d). In Fig. S9e-h, we plot the power dependence at $V_{bg} = 0.5, 0.7, 1.5$ and $1.7$V. We maintained all measurements in the linear-power regime. Note that the resistance of the Pt heater changes less than 3% from 2K to 30K; we therefore used a fixed heater voltage to maintain constant power in our temperature dependence measurements.

**Thermoelectric measurements in the quantum Hall regime**
We measured the Seebeck and Nernst voltages at $B = 12$T and $T = 2$K along an E-field linecut $E = 0.66$V/nm (Fig. S10a). The linecut was chosen inside the $1 + \nu_W$ region, in which holes are doped into the W-layer. The heater voltage was kept at $V_h = 0.1$V. The Seebeck and Nernst coefficients ($S$ and $N$) in the quantum Hall regime were well studied in the literature[11-13]. A simple picture is to write $S$ and $N$ in terms of the resistivity tensor $\hat{\rho}$ and the Peltier tensor $\hat{\alpha}$, $S = \alpha_{xx}\rho_{xx} + \alpha_{xy}\rho_{yx}$ and $N = \alpha_{xy}\rho_{xx} - \alpha_{xx}\rho_{yx}$. At half filling of the $n^{th}$ Landau level, we have $\alpha_{xx} = 0$, $\alpha_{xy} = -\frac{ek_B}{h}\ln 2$, $\rho_{yx} \approx -1/\sigma_{xy} \approx -\frac{h}{(n+1/2)e^2}$ and $\rho_{xx} \approx 0$ (in the clean limit). This yields $S = \frac{(k_B/e)\ln 2}{n+1/2} \approx \frac{59.7}{n+1/2}$ μV/K and $N = 0$. We mark the Seebeck peaks at half fillings of the Landau levels in Fig. S10a, which aligns with $N = 0$. To assign the correct Landau level indices, we determined the Chern number of the dip in $S$ (dashed line in Fig. S10b) using the Středa formula $C = \frac{n_M h}{e} \frac{dv}{dB} = 1.98$ (Ref. [14]), where $n_M = 4.39 \times 10^{12}$cm$^{-2}$ is the moiré density. In Fig. S10c, we fitted the Seebeck coefficient to $1/(n + 1/2)$ and extracted the slope to be 68.9μV/K. This is in reasonably good agreement with the theoretical value 59.7μV/K.

**Modelling the Seebeck coefficient for a single-particle band**
The WSe$_2$ band in the $2 + v_W$ region can be approximated as a single-particle parabolic band. In the Boltzmann transport theory, the Seebeck coefficient can be written as $S_{FB}(T) = -\frac{1}{eT}\frac{K_1}{K_0}$, where the kinetic coefficients $K_n$ are defined as $K_n = \int \left(-\frac{\partial f_0}{\partial E}\right)(E - \mu)^n \Sigma(E) dE$, and $\Sigma(E)$ is the generalized distribution function $\Sigma(E) = \int \tau (\hat{\mathbf{e}} \cdot \mathbf{v})^2 \delta[E(\mathbf{k}) - E] d\mathbf{k}$, where $\hat{\mathbf{e}}$ is an arbitrary unit vector for an isotropic system and $\tau$ is the relaxation time[15]. For a free-hole gas with bandwidth $W$ and energy independent relaxation time, $\Sigma(E)$ takes the form $\Sigma(E) \sim E^{3/2}$ for $-W < E < 0$. The chemical potential $\mu$ is dependent on the hole filling factor $v$ through $\mu = -W(v - 2)/2$ in the $2 + v_W$ region, in which the factor 1/2 takes into account the two-fold spin-valley degeneracy. We calculated the Seebeck coefficient map as function of $T$ and $v$ for a bandwidth $W = 13$meV in Fig. S12. The calculation predicts a larger Seebeck coefficient but qualitatively reproduces the measurement data in Fig. 3a.

## Supplementary figures

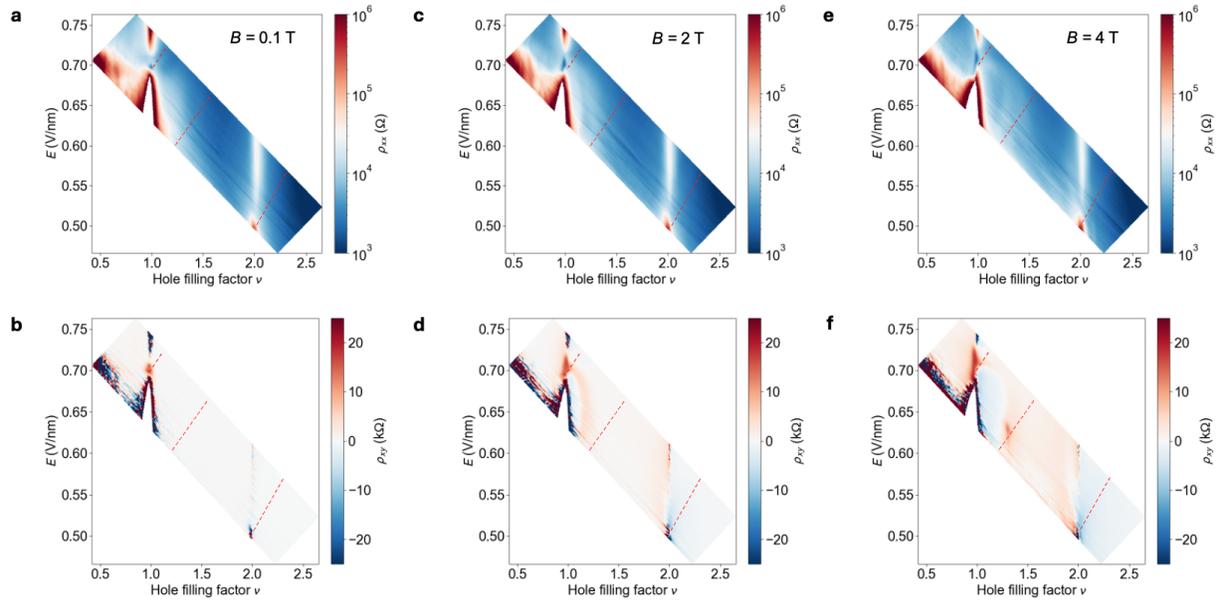

**Figure S1. $\rho_{xx}$ and $\rho_{xy}$ as a function of $\nu$ and $E$ at varying B-fields ($T = 2K$). a-f,** $\rho_{xx}$ and $\rho_{xy}$ maps at $B = 0.1\text{T}$ (**a,b**), 2T (**c,d**) and 4T (**e,f**). The phase boundaries of the $1 + \nu_W$ and $2 + \nu_W$ regions are marked by red dashed lines.

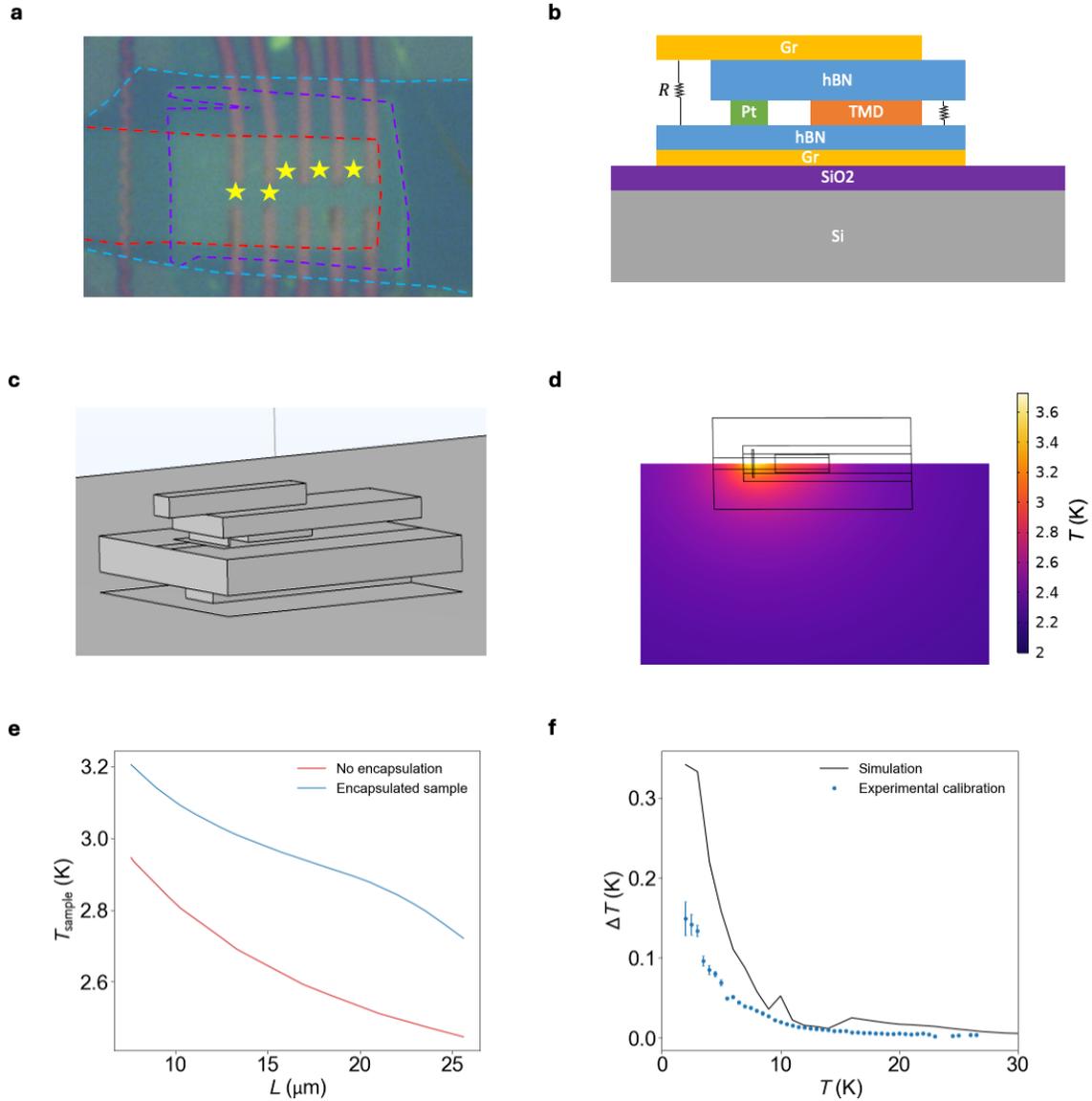

**Figure S2. Finite element analysis of the temperature gradient by COMSOL. a,** Optical micrograph of the device. The leftmost electrode is the Pt heater. The boundaries of the top graphite gate (red), bilayer device (purple) and bottom graphite gate (blue) are marked by dashed lines. Yellow stars mark the electrode pairs between which we calibrate the local temperature gradient. **b,** Schematic of the device structure. The flakes are modelled by rigid cuboids with thermal contacts. To thermally connect the bottom gate hBN to top gate hBN and graphite electrode, a thermal resistor $R$ is included. **c,** Geometry of the COMSOL model. The size of the flakes is estimated from optical micrograph (Fig. S2a). The z axis is scaled by 1000 for clarity. **d,** Steady state temperature field from simulation at $T = 2K$. A constant power $P = 4.7\mu W$ is applied to the heater. **e,** Dependence of the local temperature $T_{sample}$ on the distance $L$ to the heater extracted from the simulation. Results for an encapsulated heater and a heater-on-chip are shown in blue and red, respectively. **f,** Temperature difference $\Delta T$ between two ends of the device as a function of $T$, extracted from the simulation (solid line) and experimental calibration (dots).

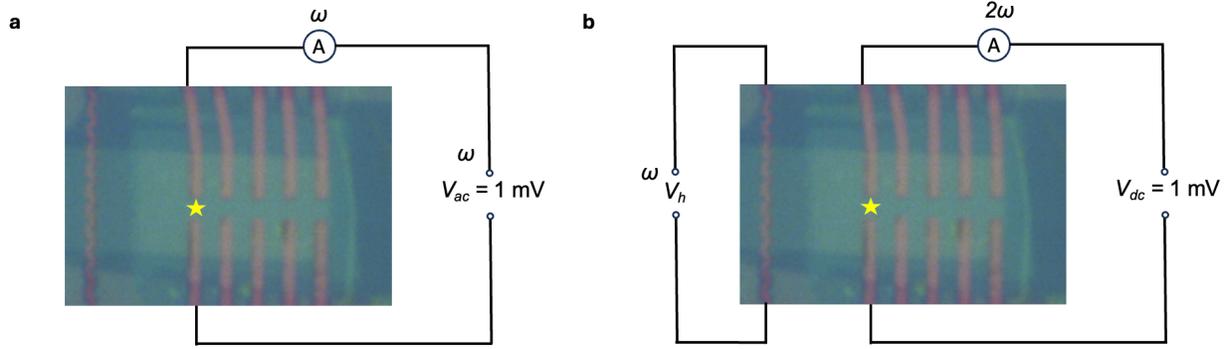

**Figure S3. Two-terminal method to calibrate local temperature rise. a,** $I$ - $T$ curve measurement. The local temperature dependence of the two-terminal resistance between two adjacent electrodes (yellow star) is served as a resistance thermometer to calibrate the local temperature rise. A small AC voltage $V_{ac}$ = 1mV ($\omega$ = 11.137Hz) is applied to the electrode pair. The temperature dependence of $I$ is measured by a lock-in amplifier. **b,** Local temperature measurement. We DC-bias the electrode pair by $V_{dc}$ = 1mV and measure the current modulation $\Delta I$ induced by an AC heater voltage $V_h$ at $2\omega$. The local temperature rise is given by $\Delta T = \frac{\Delta I}{dI/dT}$.

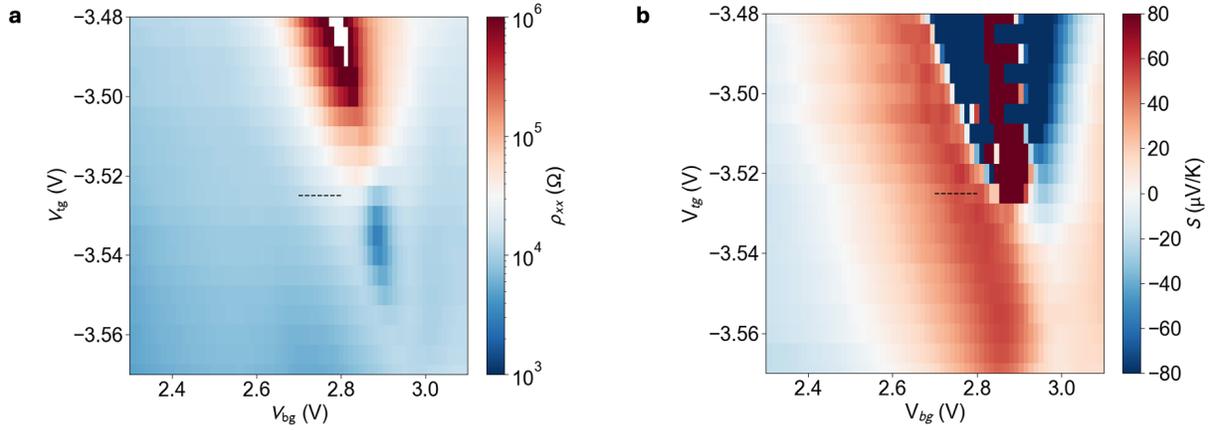

**Figure S4. Line cut used to calibrate the local temperature rise. a,** Dependence of $\rho_{xx}$ on $V_{tg}$ and $V_{bg}$ at $B = 0$T and $T = 2$K. We chose a fixed-$V_{tg}$ line cut at $V_{tg} = -3.525$V (dashed line) that is close to $\nu = 1$ to measure the $I$-$T$ curve. **b,** Dependence of the Seebeck coefficient $S$ on $V_{tg}$ and $V_{bg}$ at $B = 0$T and $T = 2$K. The dashed line marks the same line cut. Note that the calibrated temperature rise $\Delta T$ is independent of the specific line cut we chose for calibration.

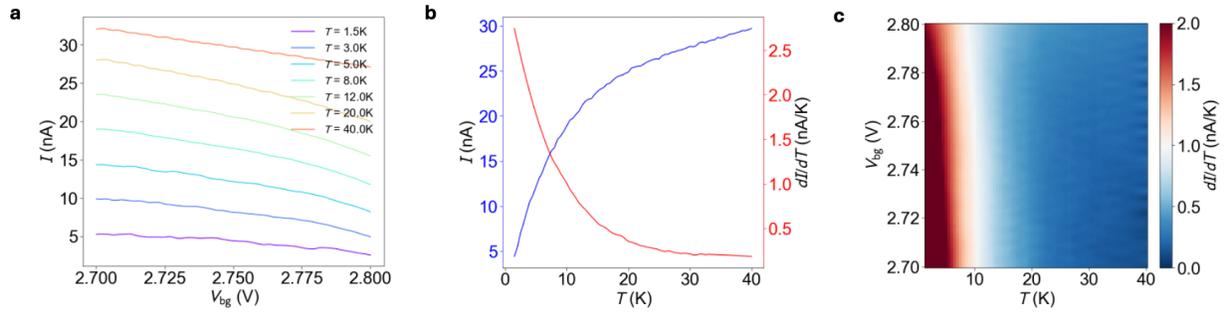

**Figure S5. Extraction of d$I$/d$T$. a,** The source-drain current $I$ (under $V_{ac}$ = 1mV) as a function of $V_{bg}$ at the fixed-$V_{tg}$ line cut in Fig. S4 at varying temperatures. **b,** Temperature dependence of $I$ at $V_{bg}$ = 2.75V (blue line) and the extracted slope d$I$/d$T$ (red line). **c,** Dependence of d$I$/d$T$ on $V_{bg}$ and $T$.

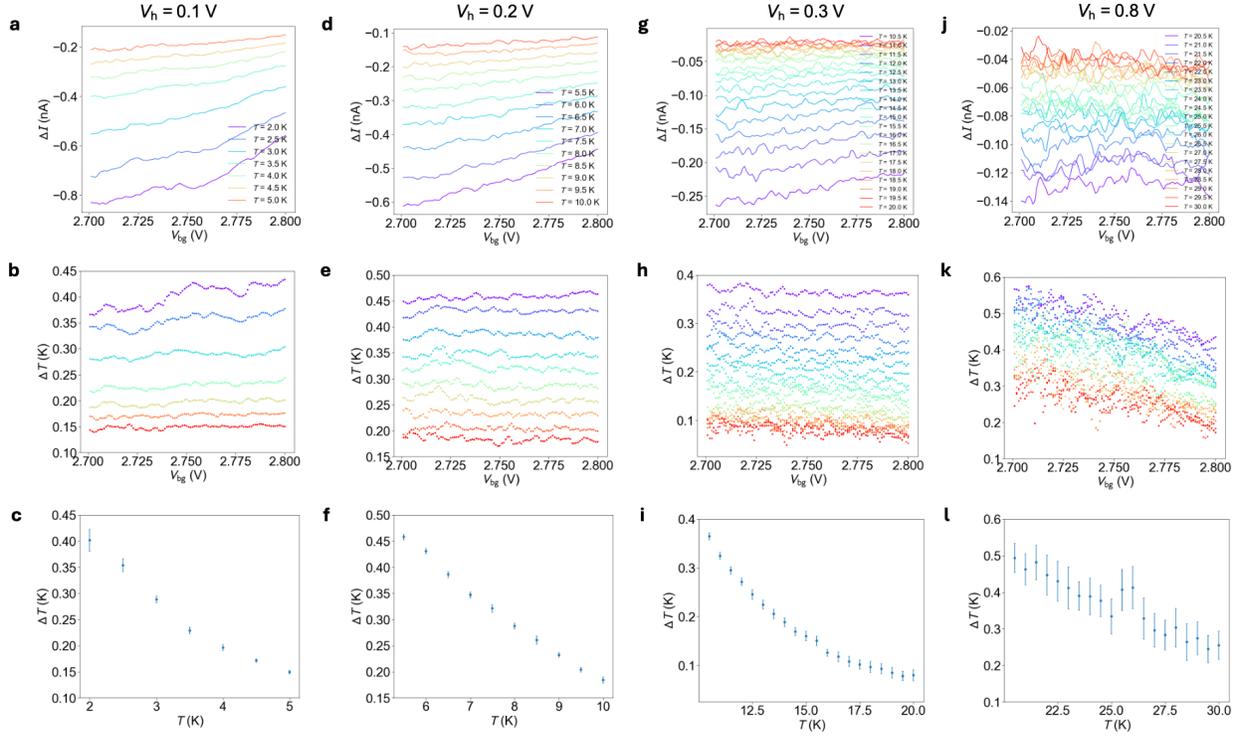

**Figure S6. Measurements of the local temperature rise $\Delta T$. a-c,** Dependence of the current modulation $\Delta I$ (**a**) and $\Delta T$ (**b**) on $V_{bg}$ at the fixed-$V_{tg}$ line cut in Fig. S4 at temperatures ranging from 2.0K to 5.0K. The AC heater bias is $V_h = 0.1$V. The local temperature rise (**c**) is obtained by averaging over $V_{bg}$ in **b**. **d-f,** Same as **a-c** for temperatures ranging from 5.5K to 10.0K with $V_h = 0.2$V. **g-i,** Same as **a-c** for temperatures ranging from 10.5K to 20.0K with $V_h = 0.3$V. **j-l,** Same as **a-c** for temperatures ranging from 20.5K to 30.0K with $V_h = 0.8$V.

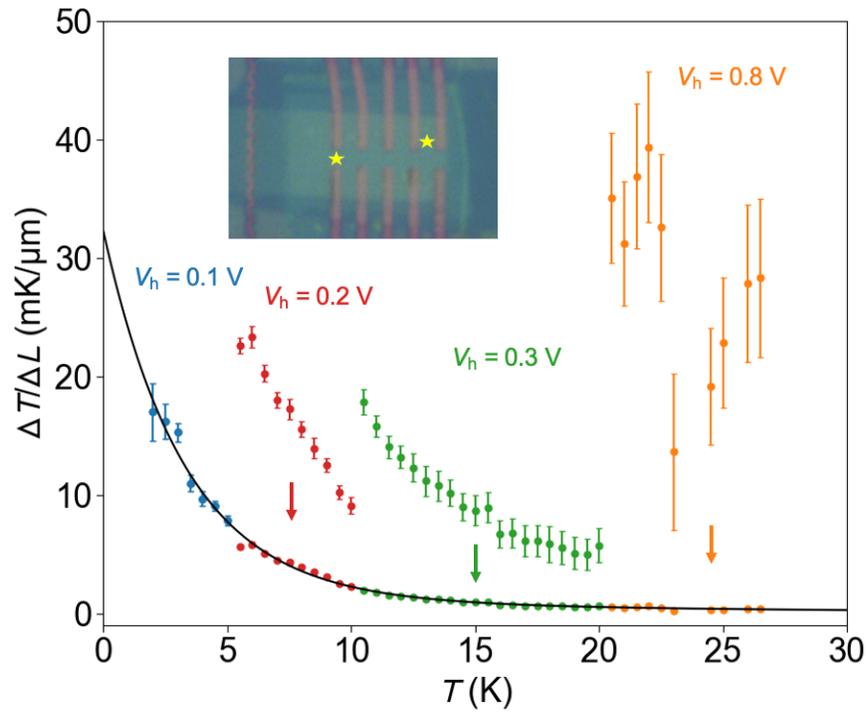

**Figure S7. Normalization of temperature gradient at different temperature windows.** The temperature gradient is normalized to $V_h = 0.1$ V and fitted to the function $Ae^{-\alpha T} + Be^{-\beta T}$ (solid line). See Methods for details. The inset shows the two points that are used to extract the temperature gradient.

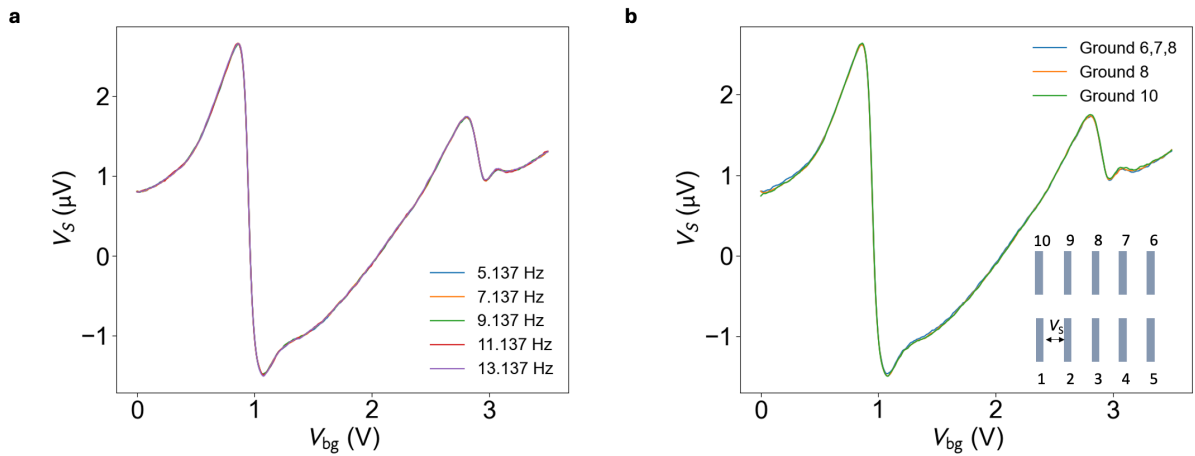

**Figure S8. Frequency and grounding tests for the thermoelectric measurements**. **a,** Dependence of the Seebeck voltage $V_S$ on $V_{bg}$ measured at $V_{tg} = -3.55$ V and at varying frequencies. **b,** Dependence of $V_S$ on $V_{bg}$ measured at $V_{tg} = -3.55$ V using different grounding electrodes. Inset: a schematic of the electrodes. $V_S$ is measured between electrodes 1 and 2.

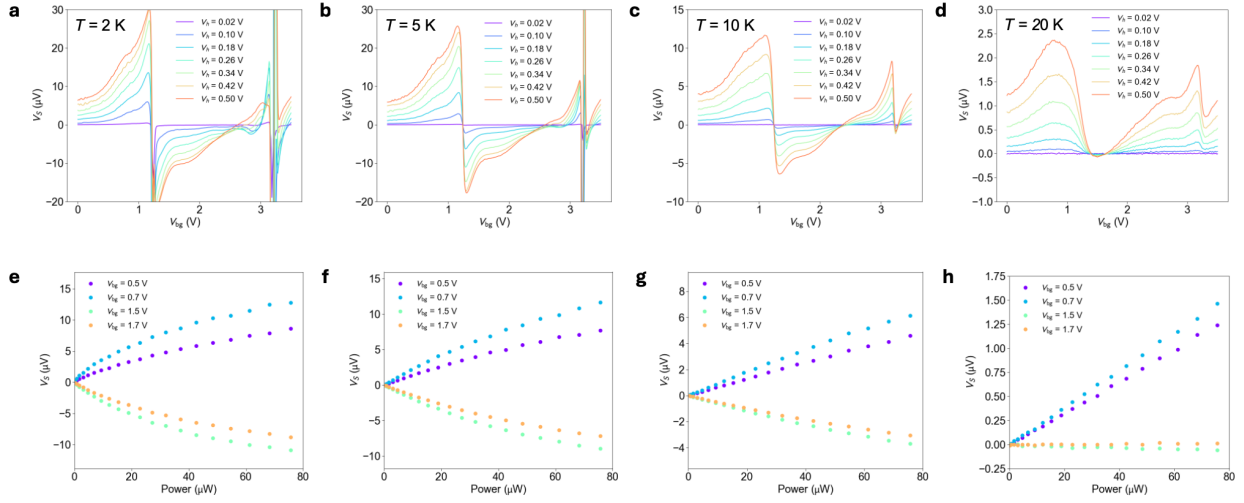

**Figure S9. Power dependence measurements. a-d,** Seebeck voltage $V_S$ as a function of $V_{bg}$ measured at $V_{tg} = -3.7$V and with varying heater bias $V_h$. The sample temperature is at $T = 2$K, 5K, 10K and 20K, respectively. **e-h**, Power dependence of $V_S$ at $T = 2$K, 5 K, 10 K and 20 K extracted from **a-d** for four different $V_{bg}$'s. The power is given by $V_h^2/R_h$, where $R_h = 2.115$kΩ is the heater resistance.

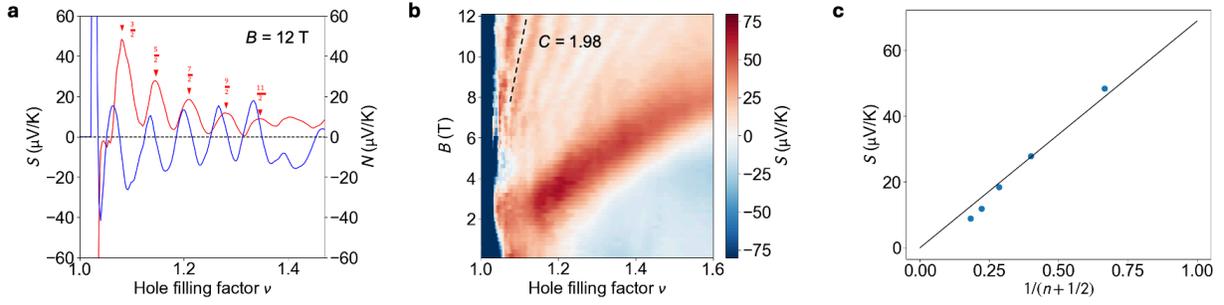

**Figure S10. Thermoelectric measurements in the quantum Hall regime. a,** Dependence of the Seebeck coefficient $S$ (red line) and the Nernst coefficient $N$ (blue line) on $\nu$ at $T = 2$K and $B = 12$T ($E = 0.66$V/nm). Red arrows label the Seebeck peaks with assignment for the Landau level fillings. **b,** Seebeck coefficient as a function of $B$ and $\nu$ at $E = 0.66$V/nm and $T = 2$K. Dashed line marks the second Landau level determined from the Středa formula $C = \frac{n_M h}{e}\frac{d\nu}{dB} = 1.98$. **c,** Linear fit of the peak value of $S$ at the half-filled Landau levels to $1/(n+1/2)$. The extracted slope is $66.7\mu$V/K, which is in good agreement with theoretical value $(k_B/e)\ln 2 \approx 59.7\mu$V/K.

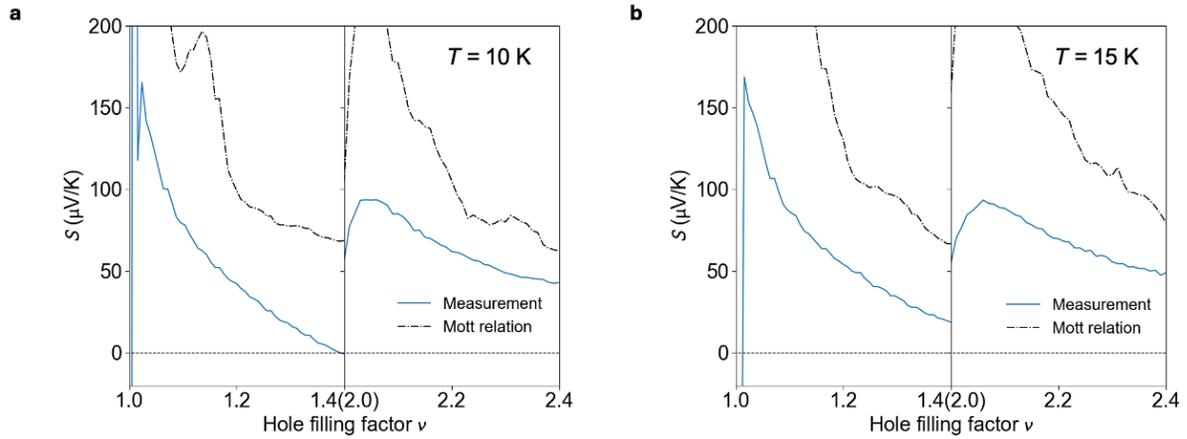

**Figure S11. Comparison with the Mott relation at elevated temperatures. a-b,** The measured Seebeck coefficient (solid lines) and the calculated $S_{\text{Mott}}$ (dashed lines) as a function of $\nu$ at $T =$ 10K, 15K, respectively. Left panel is a line cut at $E = 0.68$V/nm in the $1 + \nu_W$ region; right panel is a line cut at $E = 0.52$V/nm in the $2 + \nu_W$ region. The resistance data for the calculation of $S_{\text{Mott}}$ were measured independently at the corresponding temperatures (not shown). At elevated temperatures, $S_{\text{Mott}}$ shows the same sign as the measurement, in contrast to the different sign observed at low temperatures (see Fig. 2d). The disagreements on the absolute magnitude may result from the phonon drag effect, which is expected to set in at elevated temperatures.

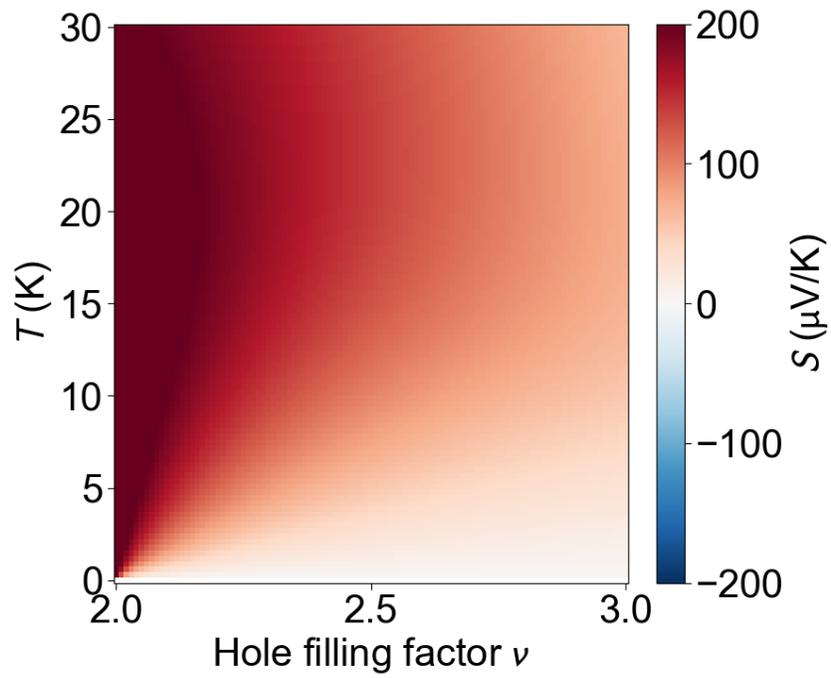

**Figure S12. Predicted Seebeck coefficient $S$ for a single-particle band as a function of $T$ and $\nu$.** The band is simplified as a truncated parabolic band with bandwidth $W = 13$meV. $S_{FB}$ was calculated using the Boltzmann transport theory with constant density of states.

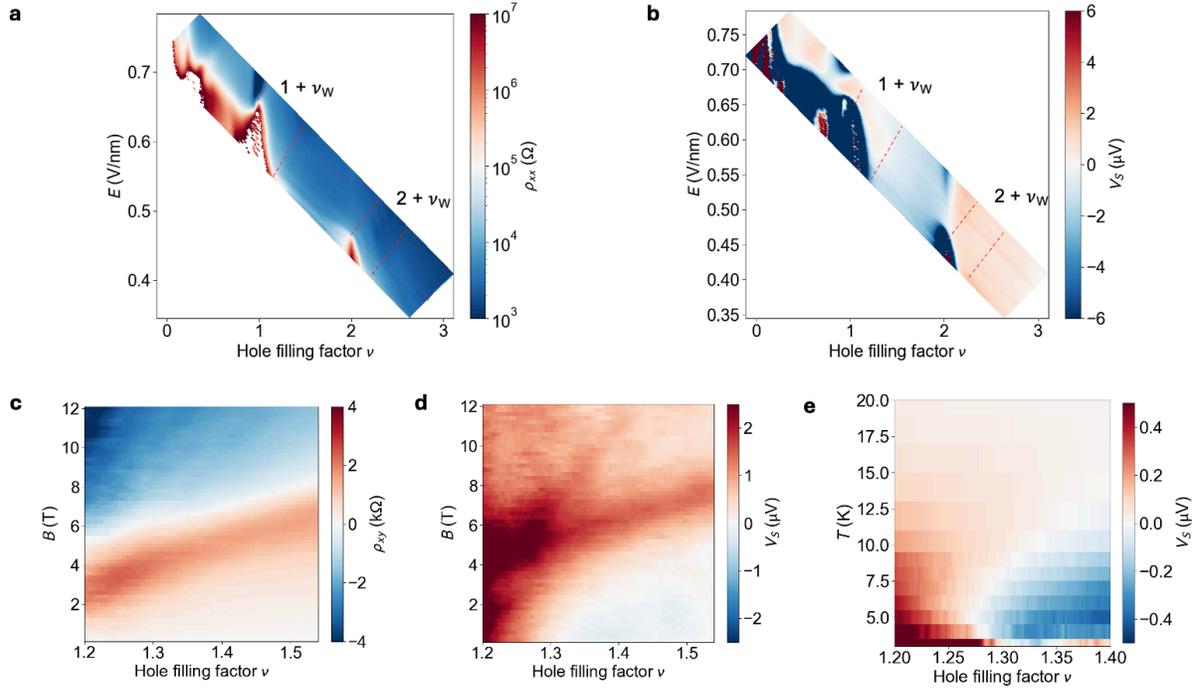

**Figure S13. Reproducibility of the main results on another angle-aligned MoTe$_2$/WSe$_2$ device.** **a,** Dependence of $\rho_{xx}$ on $E$ and $\nu$ at $T = 1.5$K and $B = 12$T. $\rho_{xx}$ is symmetrized at positive and negative $B$-fields. Dashed lines show boundaries of the $1 + \nu_W$ and $2 + \nu_W$ regions identified by the Landau levels. **b,** Dependence of the Seebeck voltage $V_S$ on $E$ and $\nu$ at $T = 5$K and $B = 0$T. **c,** $\rho_{xy}$ as a function of $B$ and $\nu$ in the $1 + \nu_W$ region ($T = 1.5$K). The magnetic Kondo breakdown characterized by a sign change in $\rho_{xy}$ is observed. **d,** The Seebeck voltage $V_S$ as a function of $B$ and $\nu$ in the $1 + \nu_W$ region ($T = 5$K). Negative Seebeck coefficient (electron-like Fermi surface) is observed for the heavy fermion phase before the magnetic Kondo breakdown. **e,** The Seebeck voltage $V_S$ as a function of $T$ and $\nu$ in the $1 + \nu_W$ region ($B = 0$T). The sign change of the Seebeck coefficient with increasing temperature shows the dissociation of Kondo singlets by thermal excitations.